\documentstyle[12pt,aaspp4]{article}

\begin{document}

\title{Thermal emission from bare quark matter surfaces of hot strange stars}

\author{Vladimir V.~Usov} 

\affil{Department of Condensed Matter Physics, Weizmann Institute, 
Rehovot 76100, Israel}

\date{Received   / Accepted  }

\begin{abstract}

We consider the thermal emission of photons and $e^+e^-$ pairs 
from the bare quark surface of a hot strange star. The radiation of 
high-energy ($\gtrsim 20$~MeV) equilibrium photons prevails at 
the surface temperature $T_{_{\rm S}}\gtrsim 5\times 10^{10}$~K, 
while below this temperature, $8\times 10^8\lesssim T_{_{\rm S}} 
< 5\times 10^{10}$~K, emission of $e^+e^-$ pairs created by the 
Coulomb barrier at the quark surface dominates. The thermal 
luminosity of a hot strange star in both photons and $e^+e^-$ 
pairs is estimated.


\end{abstract}


\section{Introduction}
Witten (\cite{W84}) pointed out that strange quark matter 
(SQM) composed 
of roughly equal numbers of up, down, and strange quarks may be the 
absolute ground state of the strong interaction, i.e., absolutely
stable with respect to $^{56}$Fe. Detailed studies showed 
that, with the uncertainties inherent in a nuclear-physics 
calculation, the existence of stable SQM 
is reasonable (e.g., Farhi \& Jaffe \cite{FJ84}; Chmaj \& Slomi\'nski
\cite{CS89}; Terazawa \cite{T90}; Kettner et al. \cite{KWWG95};
Weber \cite{W99}).
If Witten's idea is true, at least some of the compact
objects known to astronomers as pulsars, powerful, accreting X-ray
sources, X-ray and $\gamma$-ray bursters, etc. might be 
strange stars, not neutron stars as usually assumed
(Alcock, Farhi, \& Olinto \cite{AFO86}; 
Glendenning \cite{G90}, \cite{G96}; Caldwell \& Friedman \cite{CF91};
Madsen \& Olesen \cite{MO91}; Cheng \& Dai \cite{CD96}; \cite{CD98}; 
Xu, Qiao, \& Zhang \cite{XQZ99}). SQM with the density of
$\sim 5\times 10^{14}$ g~cm$^{-3}$ can exist, by hypothesis,
up to the surface of strange stars. This differs
qualitatively from the case of a neutron star surface and opens 
observational possibilities to distinguish strange stars from neutron
stars, if indeed the formers exist.
In this Letter, we consider thermal emission of photons and
$e^+e^-$ pairs from the bare SQM surfaces of hot strange stars.

\section{Bare quark surfaces of strange stars and thermal emission}
 
At the bare SQM surface of a strange star the density changes 
abruptly from $\sim 5\times 10^{14}$ g~cm$^{-3}$ to zero. The thickness
of the SQM surface is about $1~{\rm fm}=10^{-13}$~cm, which is a 
typical strong
interaction length scale. The SQM surface is a very poor radiator
of thermal photons at energies less than about 20~MeV (Alcock
et al. \cite{AFO86}). This is because the plasma frequency that is 
related to the particle  density of quarks is very high (see below).

\subsection{Thermal Equilibrium Radiation}

Hot SQM is filled with electromagnetic waves  
in thermodynamic equilibrium with quarks.  The dispersion 
relation of these waves may
be written in the following simple form (e.g., Adams, 
Ruderman, \& Woo \cite{ARW63}; Alcock et al. \cite{AFO86})

\begin{equation}
\omega^2=\omega_{\rm p}^2+k^2c^2\,,
\end{equation}

\noindent
where $\omega$ is the frequency of electromagnetic waves, $k$ is
their wavenumber, 

\begin{equation}
\omega_{\rm p}=\left({8\pi\over 3}{e^2c^2n_{\rm b}\over
\mu}\right)^{1/2}
\end{equation}

\noindent
is the plasma frequency, $n_{\rm b}$ is the baryon number density 
of SQM, and $\mu\simeq \hbar c(\pi^2n_{\rm b})^{1/3}$
is the chemical potential of quarks. 
Equation (1) is the familiar dispersion relation for a plasma, 
and its conventional interpretation may be applied to SQM. Propagating 
modes exist only for $\omega > \omega_{\rm p}$. Taking this into 
account, equation (2) yields the lower limit on the energy of 
electromagnetic quanta that are in thermodynamic equilibrium with 
quarks, $\varepsilon_\gamma =\hbar \omega > \hbar \omega_{\rm p}\simeq 
18.5(n_{\rm b}/n_0)^{1/3}$ MeV, where $n_0=0.17$~fm$^{-3}$ is normal 
nuclear matter density. At the
SQM surface where the pressure is zero, we expect
$n_{\rm b}\simeq (1.5-2)n_0$ and $\hbar \omega_{\rm p}\simeq 20-25$ MeV
(e.g., Farhi \& Jaffe \cite{FJ84}; Alcock et al. \cite{AFO86};
Chmaj, Haensel, \& Slomi\'nski
\cite{CHS91}). Therefore, the spectrum of thermal equilibrium 
photons radiated from the bare SQM surfaces of strange stars 
is expected to be very hard.  

The energy flux emitted from the unit surface of SQM in thermal 
equilibrium photons is (Bekefi \cite{B66}; Chmaj et al. \cite{CHS91})

\begin{equation}
F_{\rm eq}={\hbar \over c^2}\int_{\omega_{\rm p}}^{\infty}d\omega\,
{\omega\, (\omega^2 -\omega_{\rm p}^2)\,g(\omega )\over
{\exp\, (\hbar \omega/k_{_{\rm B}}T_{_{\rm S}})-1}}\,,
\end{equation}

\noindent
where 

\begin{equation}
g(\omega )={1\over 2\pi^2}\int_{0}^{\pi/2}d\vartheta\,
\sin\vartheta\,\cos \vartheta\, D(\omega, \vartheta)\,,
\end{equation}

\noindent
$k_{_{\rm B}}$ is the Boltzmann constant,
$D(\omega,\vartheta)$ is the coefficient of radiation transmission
from SQM to vacuum, $D=1-(R_\perp +R_\parallel)/2$, 
and

\begin{equation}
R_\perp ={\sin^2(\vartheta -\vartheta_0)\over \sin^2(\vartheta 
+ \vartheta_0)}\,,\,\,\,\,
R_\parallel={\tan^2(\vartheta -\vartheta_0)\over \tan^2(\vartheta 
+ \vartheta_0)}\,
\end{equation}

\begin{equation}
\vartheta_0= \arcsin\, [\,\sin \vartheta\,\sqrt{1-(\omega_{\rm p}/ 
\omega)^2}\,]
\end{equation}

\noindent
(see, e.g., Landau, Lifshitz, \& Pitaevskii \cite{LLP84}). 
Figure~1 shows the ratio of the equilibrium photon emissivity
of the bare surface of hot SQM to the black body
surface emissivity, $\xi_{\rm eq} = F_{\rm eq}/F_{_{\rm BB}}$, where 
$F_{_{\rm BB}}=\sigma T_{_{\rm S}}^4$, and $\sigma$ is the
Stefan-Boltzmann constant. From Figure 1 we can see that
at $T{_{\rm S}}\ll\hbar \omega_{\rm p}/k_{_{\rm B}}\sim 10^{11}$~K
the equilibrium photon radiation from the bare surface of 
a strange star is very small, compared to the black body one.

\subsection{Emission of $e^+e^-$ pairs}

It was pointed out (Usov \cite{U98}) that the bare surface of a hot
strange star may be a powerful source of $e^+e^-$ pairs which are
created in an extremely strong electric field at the quark surface and
flow away from the star. The electric field is generated because there 
are electrons with the density $n_{\rm e}
\simeq (10^{-3}-10^{-4})n_{\rm b}$ in
SQM to neutralize the electric charge of the quarks (e.g., 
Alcock et al. \cite{AFO86}). The electrons, being bound to SQM by the
electromagnetic interaction alone, are able
to move freely across the SQM surface, but clearly cannot move to
infinity because of the bulk electrostatic attraction to the quarks.
The electron distribution extends up to $\sim 10^3$ fm above the quark
surface, and a strong electric field is generated in the  
surface layer to prevent the electrons from escaping to infinity, 
counterbalancing the degeneracy  and thermal pressure. The typical 
magnitude of the electric field at the SQM surface is
$\sim 5\times 10^{17}$ V~cm$^{-1}$ (Alcock et al. \cite{AFO86}). 
This field is a few ten times higher than the critical field  
$E_{\rm cr}=m_e^2c^3/e\hbar \simeq 1.3\times 10^{16}\,\,{\rm V~cm}
^{-1}$ at which vacuum is unstable to creation of $e^+e^-$ pairs
(Schwinger \cite{S51}; Reinhardt \cite{R94}). In such a strong electric 
field, $E\gg E_{\rm cr}$, in vacuum, the pair creation rate 
is extremely high,
$W_\pm \simeq 1.7\times 10^{50}({E/ E_{\rm cr}})^2\,\,{\rm cm}^{-3}\,
{\rm s}^{-1}$. At $E\simeq 5\times 10^{17}$ V~cm$^{-1}$, we have 
$W_\pm\simeq 2.5\times 10^{53}$ 
cm$^{-3}$~s$^{-1}$. The high-electric-field region is, however, 
not a vacuum. The electrons present fill up states into which
electrons that would be created have to go. This reduces the
pair-creation rate from the vacuum value. 
It can, in fact, be shown  that at zero temperature 
the process of pair creation is suppressed altogether because 
there is no free levels for electrons to be created.

At finite temperatures, $T_{_{\rm S}}>0$, in thermodynamical 
equilibrium electronic states are only partly filled, and pair 
creation by the Coulomb barrier 
becomes possible. Since the rate of pair production when electrons 
are created into the empty states is extremely high, the empty
states below the pair creation threshold, $\varepsilon\leq 
\varepsilon_{_{\rm F}}-2m_{\rm e}c^2$, are occupied by created 
electrons almost instantly, where 
$\varepsilon_{_{\rm F}}=\hbar c(\pi^2n_{\rm e})^{1/3}$ is the Fermi 
energy of electrons, and $m_{\rm e}$ is the electron mass
(Usov \cite{U98}). Then, the rate of pair creation by
the  Coulomb barrier is determined by the process of thermalization
of electrons, which favors the empty-state production below the
pair creation threshold. The thermal energy of SQM
is the source of energy for the process of pair creation.

The flux of $e^+e^-$ pairs from the unit surface of SQM 
is (Usov \cite{U98})

\begin{equation}
\dot n_\pm\simeq \Delta r_{_E}\Delta n_e t_{\rm th}^{-1}\,,
\end{equation}

\noindent
where $\Delta r_{_E}$ is the 
thickness of the electron surface layer with a strong electric field,

\begin{equation}
\Delta n_e\simeq {3k_{_{\rm B}}T_{_{\rm S}}\over \varepsilon_{_{\rm F}}}
\exp \left(-{2m_ec^2\over k_{_{\rm B}}T_{_{\rm S}}}\right)n_e\,.
\label{Dne}
\end{equation}

\noindent 
is the density of electronic empty states with energies below the
pair creation threshold at thermodynamical equilibrium, and $t_{\rm th}$ 
is the characteristic time of thermalization of electrons.

In the surface electron layer, the spectrum of electrons
is thermalized due to electron-electron collisions, and
the thermalization time is about $ \nu_{ee}^{-1}$, 
where (e.g., Potekhin et~al. \cite{PBHY99})  

\begin{equation}
\nu_{ee}\simeq {3\over 2\pi}\left({\alpha\over \pi}\right)^{1/2}
{(k_{_{\rm B}}T_{_{\rm S}})^2\over \hbar \varepsilon_{_{\rm F}}}J(\zeta )
\end{equation}

\noindent
is the frequency of electron-electron collisions for degenerate electrons
with $\varepsilon _{_{\rm F}}\gg m_{\em e}c^2$,
$\alpha = e^2/\hbar c=1/137$ is the fine structure constant, 
(c.f. Usov \cite{U98})                                     

\begin{equation}
J(\zeta )={1\over 3}{\zeta^3\ln \,(1+2\zeta ^{-1})\over 
(1+0.074\zeta )^3}+
{\pi^5\over6}{\zeta^4\over (13.9 +\zeta)^4}\,,
\end{equation}

\begin{equation}
\zeta = 2\sqrt{{\alpha\over \pi}}{\varepsilon_{_{\rm F}}\over k_{_{\rm B}}
T_{_{\rm S}}}\simeq 0.1{\varepsilon_{_{\rm F}}\over k_{_{\rm B}}
T_{_{\rm S}}}\,.
\end{equation}

For typical parameters, $\Delta r_{_E}\simeq 500$~fm and 
$\varepsilon_{_{\rm F}}\simeq 18$~MeV, equations (7)~-~(9)
yield 

\begin{equation}
\dot n_\pm\simeq 10^{39}\left({T_{_{\rm S}}\over 10^9~{\rm K}}
\right)^3\exp \left[{-11.9\left({T_{_{\rm S}}\over 10^9~{\rm K}}
\right)^{-1}}\right]J(\zeta)\,\,{\rm s}^{-1}
\end{equation}

\noindent
within a factor of 2 or so.

The energy flux from the unit surface of SQM in $e^+e^-$
pairs created by the Coulomb barrier is $F_\pm\simeq
\varepsilon_\pm\dot n_\pm$, where $\varepsilon_\pm\simeq m_{\rm e}c^2+
k_{_{\rm B}}T_{_{\rm S}}$ is the mean energy of created particles and
$n_\pm$ is given by equation (12).  
Figure~1 shows the ratio of the SQM surface emissivity   
in $e^+e^-$ pairs to the black body surface emissivity, 
$\xi_\pm =F_\pm /F_{_{\rm BB}}$, versus the surface temperature
$T_{_{\rm S}}$. Creation of $e^+e^-$ pairs by the Coulomb
barrier is the main mechanism of thermal emission from the surface
of SQM at $8\times 10^8\lesssim T_{_{\rm S}} \lesssim 5\times 
10^{10}$~K, while the equilibrium radiation dominates at
extremely high temperatures, $T_{_{\rm S}}>5\times 10^{10}$~K. 

The SQM surface emissivity in photons at energies $\varepsilon_\gamma
< \hbar \omega_{\rm p}$ is strongly suppressed (see above). However,
low energy photons may leave SQM if they are produced by a
non-equilibrium process in the surface layer with the thickness 
of $\sim c/\omega_{\rm p}\simeq 10^{-12}$~cm. The emissivity 
of SQM in non-equilibrium  quark-quark bremsstrahlung radiation
at low energies, $\varepsilon_\gamma < \hbar \omega_{\rm p}$,
has been estimated by Chmaj et al. (\cite{CHS91}), $\xi _{\rm neq}=
F_{\rm neq}/F_{_{\rm BB}}\simeq 10^{-4}$. Chmaj et al. (\cite{CHS91})
made a few assumptions that led to over-estimate of $\xi_{\rm neq}$, 
and $\xi_{\rm neq}\simeq 10^{-4}$ is, in fact, an upper limit.
Hence, our estimate of the SQM surface 
emissivity, $\xi =\xi_{\rm eq}+\xi_\pm$, is valid for 
$T_{_{\rm S}}\gtrsim 8\times 10^8$~K when $\xi_\pm$ 
is more than $\sim 10^{-4}$ (see Fig.~1). For 
$T_{_{\rm S}} < 8\times 10^8$~K the value of $\xi$ is  
uncertain, $\xi_\pm \lesssim\xi \lesssim 10^{-4}$, especially 
at low energies of photons.
 
\subsection{Thermal radiation from a hot strange star}

At $T_{_{\rm S}}\gtrsim 8\times 10^8$~K  
the total luminosity of a bare strange star  
in both photons and $e^+e^-$ pairs is

\begin{equation}
L=L_{\rm eq}+L_\pm = 4\pi R^2 (F_{\rm eq} +F_\pm )\,,
\end{equation} 

\noindent
where $R\simeq 10^6$~cm is the radius of the strange star. Figure~2 
shows the value of $L$ as a function of the surface temperature 
$T_{_{\rm S}}$.  At $T_{_{\rm S}} < 8\times 
10^8$~K the total luminosity is somewhere between $\sim 4\pi R^2F_\pm$ 
and $\sim 10^{-3}R^2F_{_{\rm BB}}$.

At $T_{_{\rm S}} \gtrsim 8\times 10^8$~K the luminosity 
in $e^+e^-$ pairs created by the Coulomb barrier at the SQM surface 
is very high,  $L_\pm \gtrsim 10^{40}$ ergs~s$^{-1}$ (see Fig.~2), 
that is at least four orders of magnitude higher than 

\begin{equation}
L_\pm ^{\rm max}\simeq 4\pi m_{\rm e}c^3R/\sigma_{_{\rm T}}\simeq
10^{36}\,\,{\rm ergs~s}^{-1}\,,
\end{equation}

\noindent
where $\sigma_{_{\rm T}}$ is the Thomson cross-section.
In this case, the time-scale $t_{\rm ann}\sim (n_\pm
\sigma_{_{\rm T}}c)^{-1}$ for annihilation of $e^+e^-$
pairs is much shorter than the time-scale $t_{\rm esc}\sim
R/c$ for their escape, $t_{\rm ann}/t_{\rm esc}\simeq 
L^{\rm max}_\pm/L_\pm <10^{-4}\ll 1$,
and $e^+e^-$ pairs outflowing from
the stellar surface mostly annihilate in the vicinity 
of the strange star, $r\sim R$ (Beloborodov \cite{B99} and
references therein). If the total
luminosity is not very high, $L\lesssim 10^{42}~{\rm ergs~s}^{-1}$, 
the luminosity in $e^+e^-$ pairs at the distance $r\gg R$
cannot be essentially more than $L_\pm^{\rm max}$.
At $L\gg 10^{42}~{\rm ergs~s}^{-1}$ the outflowing $e^+e^-$ wind
is relativistic, and the luminosity in pairs at $r\gg R$ is
$\sim L_\pm^{\rm max}(L/10^{42}~{\rm ergs~s}^{-1})^{3/4}
< 10^{-6}L$ (e.g., Usov \cite{U94}; Lyutikov~\& Usov \cite{LU00}).
Hence, far from a bare strange star with the surface temperature
$T_{_{\rm S}}\gtrsim 8\times 10^8$~K the photon luminosity 
dominates irrespective of $T_{_{\rm S}}$ and practically 
coincides with the total luminosity given by equation (13). 

\section{Discussion}

In this paper, we have considered the thermal emission from
the bare SQM surface of a hot strange star. The SQM surface emissivity  
in both photons and $e^+e^-$ pairs is estimated for 
$T_{_{\rm S}}\gtrsim 8\times 10^8$~K. We found that at $T_{_{\rm S}}
\gtrsim 1.5\times 10^9$~K the SQM surface emissivity is $\gtrsim 
10$\% of the black body surface emissivity. Below this temperature, 
$T_{_{\rm S}}\lesssim 1.5\times 10^9$~K, the SQM surface emissivity 
decreases rapidly with decrease of $T_{_{\rm S}}$, and 
at $T_{_{\rm S}} < 8\times 10^8$~K it is at least
four orders of magnitude smaller than the emissivity of the black
body surface.

Recently, it was argured that SQM is a superconductor if its
temperature is not too high (for a review, see
Rajagopal \& Wilczek \cite{RW00}). In this case, the non-equilibrium
quark-quark bremsstrahlung radiation discussed in (Chmai et al. 
\cite{CHS91}) is completely suppressed, and equation (13) may be used 
at $T_{_{\rm S}} < 8\times 10^8$~K as well.  

"Normal" matter (ions and electrons) may be present at the SQM surface 
of a strange star. The ions in the inner layer are supported against 
the gravitational attraction to the underlying strange star 
by the very strong electric field of the Coulomb barrier.
There is an upper limit to the amount of normal matter at the quark 
surface, $\Delta M\lesssim 10^{-5}M_\odot$ 
(Glendenning \& Weber \cite{GW92}). Such 
a massive envelope of normal matter with $\Delta M\sim 10^{-5}M_\odot$
completely obscures the quark surface. However, it was pointed out 
(e.g., Haensel, Paczy\'nski,  \& Amsterdamski
\cite{HPA}) that a strange star at the moment of its formation is very 
hot. The temperature in the stellar interior may be as high as a 
few $\times 10^{11}$ K. The rate of mass ejection from such a 
hot compact star is very high (e.g., Woosley \& Baron \cite{WB92}; 
Levinson \& Eichler \cite{LE93}; Woosley \cite{W93}), and most
probably, the normal-matter envelope is blown away by radiation 
pressure. High temperatures also lead to considerable reduction of 
the Coulomb barrier, increasing the tunneling of nuclei through the 
barrier toward the SQM surface (Kettner et al.
\cite{KWWG95}). Therefore, it is natural to expect the 
SQM surface of a very young strange star to be nearly (or completely) 
bare. It was argued that the normal-matter atmosphere of such a  
star remains optically thin until the temperature of the SQM surface 
is higher than $\sim 3\times 10^7$ K (Usov \cite{U97}).

Since SQM at the surface of a bare strange star is bound via strong
interaction rather than gravity, such a star can radiate
at the luminosity greatly exceeding the Eddington limit at the stellar
mass of $\sim M_\odot$ (Alcock~et~al.
\cite{AFO86}; Chmaj~et~al. \cite{CHS91}; Usov \cite{U98} and this
paper). Therefore, bare strange stars are reasonable candidates
for soft $\gamma$-ray repeaters (SGRs) that are the sources of short 
bursts of high-frequency (X-ray and $\gamma$-ray) emission with 
Super-Eddington luminosities, up to $\sim 10^{43}-10^{45}~{\rm ergs~s}
^{-1}$. The bursting activity of SGRs may be explained by fast heating 
of the stellar surface up to the temperature of 
$\sim (1-2)\times 10^9$~K (see Fig.~2) and its subsequent 
thermal emission. The heating mechanism may be either 
impacts of comets onto bare strange stars (Zhang, Xu, \& Qiao 
\cite{ZXQ00}) or fast decay of superstrong ($\sim 10^{14}-
10^{15}$~G) magnetic fields (Usov \cite{U84}; 
Thompson \& Duncan \cite{TD95}; Heyl \& Kulkarni \cite{HK98}).

\begin{acknowledgements} 
This work was supported by the Israel Science Foundation of
the Israel Academy of Sciences and Humanities.
\end{acknowledgements}

\onecolumn

FIGURE CAPTIONS

Fig. 1. Total emissivity of bare SQM surface
(solid line), which is the sum of emissivities 
in equilibrium photons (dashed line) and $e^+e^-$ pairs 
(dotted line), divided by the black body emissivity,
$\xi=\xi_{\rm eq}+\xi_\pm$.
Our calculations are valid for the surface temperature
$T_{_{\rm S}}\gtrsim 8\times 10^8$~K when $\xi$ is more than
the upper limit on $\xi_{\rm neq}$, $\xi_{\rm neq}\lesssim 10^{-4}$,
which is shown by dot-dashed line. 

Fig. 2. The total luminosity of a bare strange star (solid line)
as a function of the surface temperature $T_{_{\rm S}}$,
$L=L_{\rm eq}+L_\pm$, where $L_{\rm eq}$ and $L_\pm$ are the
luminosities in thermal equilibrium photons (dashed line)
and $e^+e^-$ pairs (dotted line) respectively. The upper limit 
on the luminosity in non-equilibrium photons, 
$L_{\rm neq}\lesssim 10^{-4}4\pi R^2F_{_{\rm BB}}$, is shown 
by dot-dashed line.

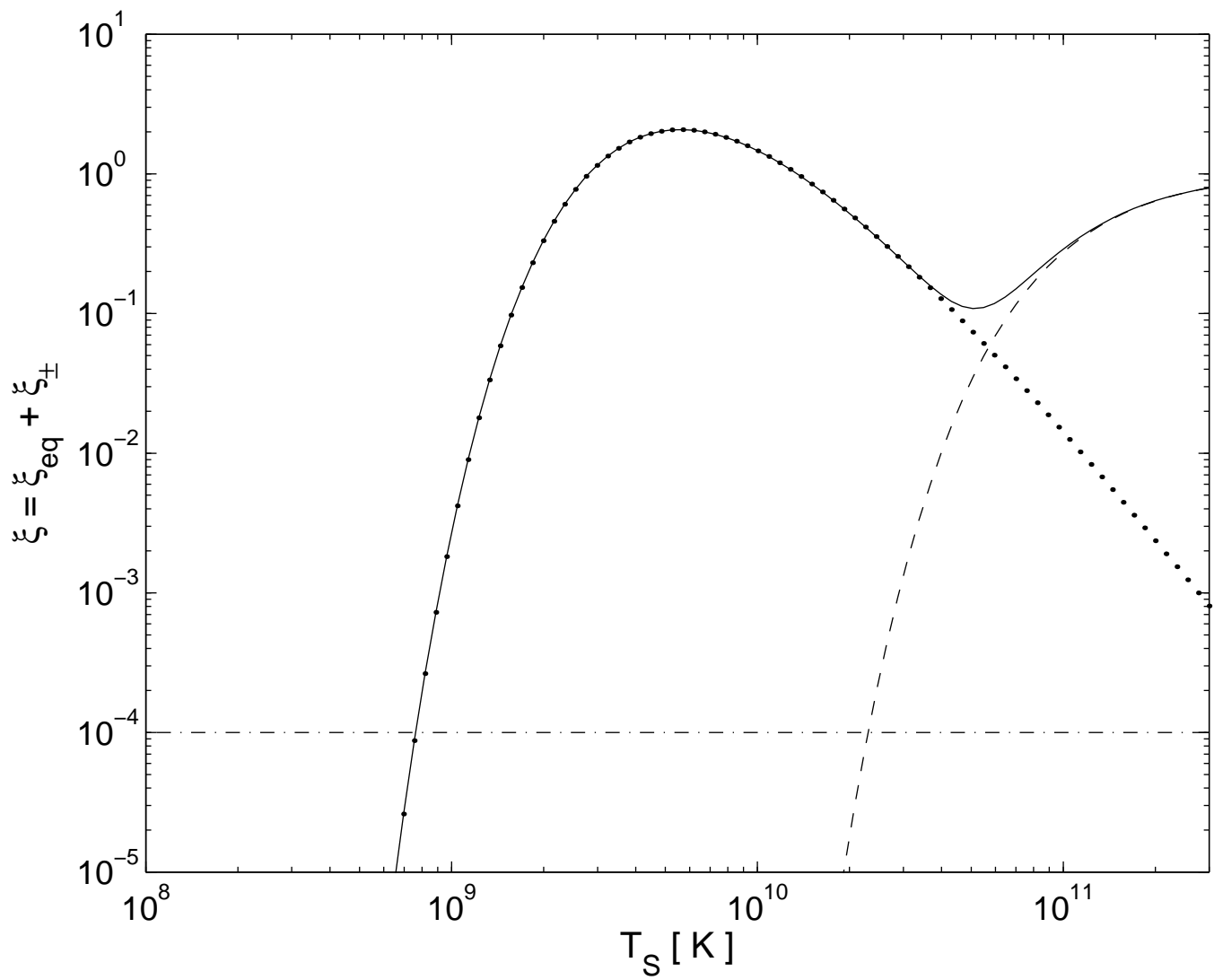
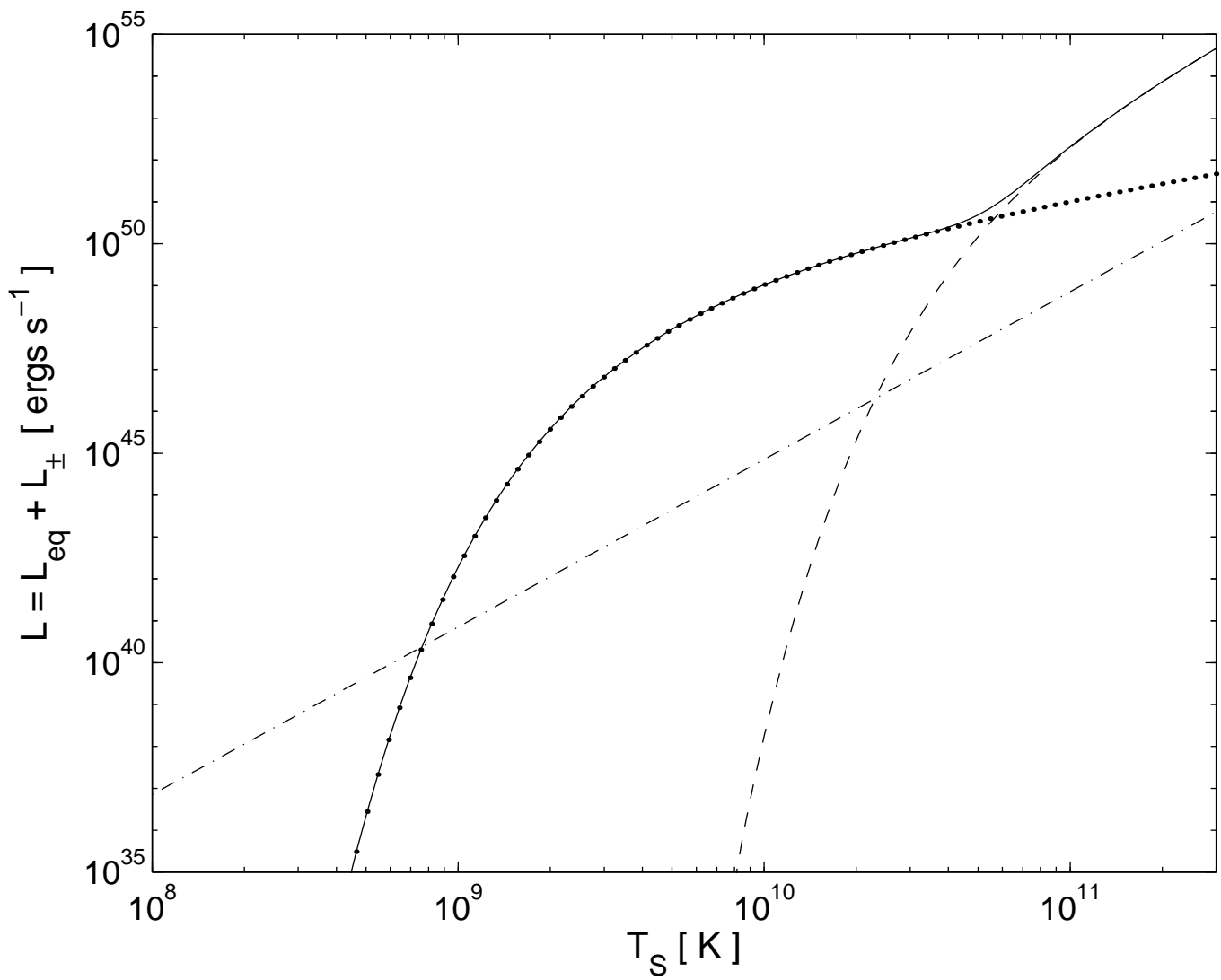
\end{document}